\documentstyle[aps,prb,preprint]{revtex}
\begin{document}
\draft
\def\dfrac#1#2{{\displaystyle{#1\over#2}}}
\title{Phase diagram regions deduced for strongly correlated
systems via unitary transformation}
\author{Endre~Kov\'acs and Zsolt~Gul\'{a}csi}

\address{Department of Theoretical Physics, Lajos Kossuth University, H-4010
Debrecen, Hungary }
\date{December, 1999}
\maketitle
\begin{abstract}
From known phase diagram regions of different model Hamiltonians describing
strongly correlated systems we deduced new domains of the ground state phase
diagram of the same model by an unitary transformation. Different types of
extended Hubbard Hamiltonians were used for the starting point and the
existence of new stable spin-density wave, charge-density wave, ferromagnetic
state and a paramagnetic insulator is demonstrated. The used procedure itself
is dimension independent.
\end{abstract}

\newpage

\section{Introduction}

Because of their interesting properties, the strongly correlated systems
polarized a huge effort in last decades given in specially by developments
in high critical temperature superconductivity (Degiorgi 1999), heavy-fermion
systems (Mathur et al. 1998), metal-insulator transition and
layered materials (de Boer et al. 1999). The rigorous results
obtained in this field are related mainly to one dimensional case being
deduced by Bethe anzats and exact result for $D \: > \: 1$ are extremely rare.
Generated by these state of facts, in this paper an
interesting procedure is presented that allows the construction
of new exact ground states (GS) in
arbitrary dimensions for strongly correlated systems starting from known
domains of the phase diagram of a given model in the same dimension. The
essence of the method is simple. Given an examined Hamiltonian ($H$) together
with its ground state $\psi_g$ valid in a restricted region of the phase
diagram $R$, based on an unitary transformation $U U^{\dagger} \: = \: 1$ we
obtain $H' \: = \: U \: H \: U^{\dagger}, \: \psi_g' \: = \: U \: \psi_g$. The
eigenvalues being unmodified by an unitary transformation, $\psi_g'$
represents the GS wave function for $H'$. For a class of analysed
Hamiltonians, $H$ can be given however with a coupling constant set
$\{ \: g_i \: \}$ and an operatorial terms set $\{ \: O_i \: \}$ as
$H \: = \: \sum_i \: g_i \: O_i$. Taking $U$ in such a way to transform
the components $O_i$ within the $\{ \: O_i \: \}$ set (i.e. $U \: O_{i_1} \:
= \: O_{i_2} \: )$, the unitary transformation change in fact
$\{ \: g_i \: \}$ to $\{ \: g_i' \: \}$. Based on this observation, from $R$,
the transformed $R'$ region can be deduced, $R'$ being the parameter space
region where $\psi_g'$ is valid.

In this paper we are using an explicitely given $U$ unitary transformation
that transforms a spin density wave (SDW) into a charge density wave (CDW)
wave function, a superconductor into a ferromagnet, a phase separation in
spin into a phase separation in charge, and vice versa. The $\{ \: O_i \: \}$
operatorial set is so fixed to preserve the analysed model in the class of
extended Hubbard models. The concretely analysed cases are connected to the
model used by de Boer at al. (1995 a, b) and Montorsi and Campbell (1996)
describeing superconducting properties, and Strack and Vollhardt (1993) and
Aligia et al. (1995) studying different density wave phases. Our results
underline the presence of new ferromagnetic, SDW and insulating GS wave
functions in the phase diagram of the enumerated models.

The remaining part of the paper is conscructed as follows: in Capter II.
we describe the family of the model Hamiltonians that we analyse and define
our unitary transformation $U$ used during the paper. Chapter III. contains
the study of different concrete Hamiltonians and the results obtained. In
Capter IV. we present the summary, and Appendices containing the mathematical
details close the paper.

\section{The system and method used}

For the starting point we are introducing below a class of model
Hamiltonians that are used in the remaining part of the paper for the
exemplification of the used procedure.

\subsection{The studied class of models.}

We are developing here our study in a class of extended Hubbard models that
generically can be given in the following form
\begin{eqnarray}
&&\hat{H} \: = \: - \: t \: \hat{t} \: + \: U \: \hat{U} \: + \: X \: \hat{X}
\: + \: V \: \hat{V} \: + \: J_{z} \: \hat{J}_{z} \: + \: \frac{J_{xy}}{2} \:
\hat{J}_{xy} \: + \: Y \: \hat{Y} \: + \: R \: \hat{R}
\nonumber \\
&& \quad + \: P \: \hat{P} \: + \: Q \: \hat{Q} \: + \: \mu \: \sum_{j=1}^{N}
\: n_{j} \: + \: h \: \sum_{j=1}^{N} \: ( \: n_{j\uparrow} \: - \:
n_{j\downarrow} \: ) \: .
\label{eqH}
\end{eqnarray}
In this expression a general $A \: \hat {A}$ term represent the contribution
of the $\hat{A}$ operator in the extended Hubbard Hamiltonian $H$, taken into
account with the coupling constant $A$. The terms taken into consideration,
in order, are the following.

The kinetic energy contribution has the form
\begin{eqnarray}
\hat{t} \: = \: \sum_{<j,i>,\sigma} \: ( \: c^{\dagger}_{j\sigma} \:
c_{i\sigma} \: + \: c_{i\sigma}^{\dagger} \: c_{j\sigma} \: ) \: ,
\end{eqnarray}
where $c_{i\sigma}$ are canonical Fermi operators wich describe
electrons on a d-dimensional lattice, $<j,i>$ denoting nearest-neighbor
sites. The Hubbard on-site interaction is taken as
\begin{eqnarray}
U \hat{U} \: = \: U \: \sum^{N}_{j=1} \: ( \: n_{j\uparrow} \: - \:
\frac{1}{2} \: ) \: ( \: n_{j\downarrow} \: - \: \frac{1}{2} \: ) \: ,
\label{U}
\end{eqnarray}
where $U$ represents the Hubbard coupling and
$ \: n_{i\sigma} \: = \: c_{i\sigma}^{\dagger} \: c_{i\sigma} \: $
denotes the particle number operator for electrons with spin $\sigma$ on
site $i$. The bond-charge interaction is taken as
\begin{eqnarray}
\hat{X} \: = \: \sum_{<i,j>,\sigma} \: ( \: c^{\dagger}_{j\sigma} \:
c_{i\sigma} \: + \: c_{i\sigma}^{\dagger} \: c_{j\sigma} \: ) \: ( \:
n_{j,-\sigma} \: + \: n_{i,-\sigma} \: ) \: .
\end{eqnarray}
For a density-density type interaction we have taken into account the
\begin{eqnarray}
\hat{V} \: = \: \sum_{<j,i>} \: ( \: n_{j} \: - \: 1 \: ) \: ( \: n_{i}
\: - \: 1 \: ) \: ,
\end{eqnarray}
contribution that describes a nearest-neighbor Coulomb interaction.

Heisenberg type spin-spin interactions follows written for spin densities.
These interactions have been taken anisotropic and are represented by
\begin{eqnarray}
&& \hat{J_{z}} \: = \: \sum_{<j,i>} \: S^{z}_{j} \: S^{z}_{i}
\\
&& \hat{J_{xy}} \: = \: \sum_{<j,i>} \: ( \: S_{j}^{+} \: S_{i}^{-} \: +
\: S_{j}^{-} \: S_{i}^{+} \: ) \: .
\label{heis}
\end{eqnarray}
The spin operators in this expression are standard, namely
$ \: S^{z}_{j} \: = \: 1 / 2 \: ( \: n_{j\uparrow} \: - \: n_{j\downarrow}
\: ) \: , \quad  S_{j}^{+} \: = \: c_{j\uparrow}^{\dagger} \: c_{j\downarrow}
\: , \quad S_{j}^{-} \: = \: c_{j\downarrow}^{\dagger} \: c_{j\uparrow} \:
$ .

A pair hopping term follows given by the Hamiltonian term
\begin{eqnarray}
\hat{Y} \: = \: \sum_{<j,i>} \: ( \: c_{j\uparrow}^{\dagger} \:
c_{j\downarrow}^{\dagger} \: c_{i\downarrow} \: c_{i\uparrow} \: + \:
c_{i\uparrow}^{\dagger} \: c_{i\downarrow}^{\dagger} \: c_{j\downarrow}
\: c_{j\uparrow} \: ) \: .
\end{eqnarray}
After this step three-particle interactions are taken into consideration. A
correlated hopping term testing the opposite spin occupancy along the bond of
the hopping is represented by
\begin{eqnarray}
\hat{R} \: = \: \sum_{<j,i>,\sigma} \: ( \: c^{\dagger}_{j\sigma} \:
c_{i\sigma} \: + \: c_{i\sigma}^{\dagger} \: c_{j\sigma} \: ) \:
n_{j,-\sigma} \: n_{i,-\sigma} \: .
\end{eqnarray}
Correlation effects produced by an empty site or a double occupancy are
described by
\begin{eqnarray}
\hat{P} \: = \: \sum_{<j,i>} \: [ \: ( \: n_{j\uparrow} \: - \: \frac{1}{2}
\: ) \: ( \: n_{j\downarrow} \: - \: \frac{1}{2} \: ) \: ( \: n_{i} \: -
\: 1 \: ) \: + \: ( \: n_{i\uparrow} \: - \: \frac{1}{2} \: ) \: ( \:
n_{i\downarrow} \: - \: \frac{1}{2} \: ) \: ( \: n_{j} \: - \: 1 \: )
\: ] \: .
\end{eqnarray}
Finally, a four-particle interaction takes into consideration a double
occupancy - double occupancy type correlation effect along different bonds
\begin{eqnarray}
\hat{Q} \: = \: \sum_{<j,i>} \: ( \: n_{j\uparrow} \: - \: \frac{1}{2} \:
) \: ( \: n_{j\downarrow} \: - \: \frac{1}{2} \: ) \: ( \: n_{i\uparrow}
\: - \: \frac{1}{2} \: ) \: ( \: n_{i\downarrow} \: - \: \frac{1}{2}
\: ) \: .
\end{eqnarray}
Furthermore, in the starting Hamiltonian given in Eq.(\ref{eqH}), $\mu$
represents the chemical potencial, and $h$ is the external magnetic field
applied in $z$ direction.

Concerning the notations from the Hamiltonian $\hat{H}$ we mention some
limiting cases that emerges during the presentation. When the Heisenberg type
interaction Eq.(\ref{heis}) becomes isotropic, the notation of the coupling
constant becomes $W \: = \: J_z \: = J_{xy}$ and the corresponding
Hamiltonian term is denoted by
\begin{eqnarray}
\hat{W} \: = \: \sum_{<j,k>} \: \vec{S}_{j} \: \vec{S}_{k} \: ,
\end{eqnarray}
where we have $ \vec{S}_{j} \: \vec{S}_{k} \: = \: S_{j}^{z} \: S_{k}^{z} \:
+ \: 1/2 \: ( \: S_{j}^{+} \: S_{k}^{-} \: + \: S_{j}^{-} \: S_{k}^{+} \: )$.
Furthermore, in case of $ \: V \: = \: Y \: $, the common notation of the
$V \hat{V} + Y \hat{Y}$ interactions becomes $S \hat{S}$ where
\begin{eqnarray}
\hat{S} \: = \: \sum_{<j,k>} \: \vec{S'}_{j} \: \vec{S'}_{k} \: = \:
\sum_{<j,k>} \: \left[ \: S_{j}^{'z} \: S_{k}^{'z} \: + \: \frac{1}{2} \:
( \: S_{j}^{'+} \: S_{k}^{'-} \: + \: S_{j}^{'-} \: S_{k}^{'+} \: ) \:
\right] \: ,
\end{eqnarray}
where
$S_{j}^{'z} \: = \: \frac{1}{2} \: ( \: n_{j\uparrow} \: + \: n_{j\downarrow}
\: - \: 1 \: ) \: , \quad S_{j}^{'+} \: = \: c^{\dagger}_{j\uparrow} \:
c_{j\downarrow}^{\dagger} \: , \quad S_{j}^{'-} \: = \: c_{j\downarrow} \:
c_{j\uparrow} \: $ .

In the following paragraph we are presenting some basic wave functions that
emerge during the paper.

\subsection{The applied wave functions.}

We are presenting in this Section the ground state wave functions used
during the paper.

The superconductor state is considered in the form of an $\eta$-pairing state
as follows
\begin{eqnarray}
| \: \Psi_{\eta} \: \rangle \: = \: ( \: S_{+} \: )^{N_{\eta}} \:
| \: 0 \: \rangle \: ,
\end{eqnarray}
where $| \: 0 \: \rangle$ is the bare vacuum with no fermions present, and
\begin{eqnarray}
S_{+} \: = \: \sum^{N}_{i=1} \: e^{i P_i} \: c_{i\uparrow}^{\dagger} \:
c_{i\downarrow}^{\dagger} \: ,
\end{eqnarray}
creates the $\eta$-pairs within the sytem. Here $N$ represents the number of
lattice sites, and $N_{\eta}$ the number of $\eta$-pairs present.

The ferromagnetic wave-function is taken as
\begin{eqnarray}
| \: \Psi_{F} \: \rangle \: = \: \prod_{i \in \cal{B}} \:
\prod_{j\in \cal{B'}} \: c^{\dagger}_{i\uparrow} \: c^{\dagger}_{j\downarrow}
\: | \: 0 \: \rangle \: ,
\end{eqnarray}
where $\cal{B}$ and  $\cal{B'}$ are arbitrary disjoint sets of lattice
points, wich together built up the whole lattice. The number of components in
$\cal{B}$ and $\cal{B}'$ is considered different.

The density wave functions considered are the following. The
spin-density-wave (SDW) state is given by
\begin{eqnarray}
| \: \Psi_{S} \: \rangle \: = \: \prod_{i\in \cal{A}} \:
\prod_{j\in \cal{A'}} \: c^{\dagger}_{i\uparrow} \: c^{\dagger}_{j\downarrow}
\: | \: 0 \: \rangle \: ,
\label{SDW}
\end{eqnarray}
where $\cal{A}$ and $\cal{A}'$ represent sublattices of a given lattice.
We mention that during the paper, in case of density waves we are working on
bipartite lattice with odd and even sublattices.

In case of charge-density wave (CDW) we have
\begin{eqnarray}
| \: \Psi_{C} \: \rangle \: = \: \prod_{i\in \cal{A}} \:
c^{\dagger}_{i\uparrow} \: c^{\dagger}_{i\downarrow} \:
| \: 0 \: \rangle \: ,
\end{eqnarray}
where the ${\cal{A}}$ has the meaning of a sublattice.

\subsection{The unitary transformation used}

The unitary transformation that we analysed during the paper has the form
\begin{eqnarray}
U \: = \: \prod_{i=1}^{N} \: ( \: e^{i P_i} \: c^{\dagger}_{i\downarrow} \:
+ \: c_{i\downarrow} \: ) \: .
\label{uni}
\end{eqnarray}
As presented in Appendix 1., $U$ given in Eq.(\ref{uni}) represents indeed an
unitary transformation. It transforms the presented wave functions
(see Appendix 2. for details) as follows
\begin{eqnarray}
&& U \: | \: \Psi_{\eta} \: \rangle \: = \: | \: \Psi_{F} \: \rangle \: ,
\nonumber\\
&& U \: | \: \Psi_{S} \: \rangle \: = \: | \: \Psi_{C} \: \rangle \: .
\label{x1}
\end{eqnarray}
Based on Eq.(\ref{uni}) the Hamiltonian from Eq.(\ref{eqH}) can be unitary
transformed. This procedure may be started with the transformation of the
creator and annihilator operators (details presented in Appendix 3.).
\begin{eqnarray}
U \: c_{i\uparrow} \: U^{\dagger} & \: = \: & (-1)^{N} \: c_{i\uparrow} \: ,
\nonumber \\
U \: c_{i\uparrow}^{\dagger} \: U^{\dagger} \: & = & \: (-1)^{N} \:
c_{i\uparrow}^{\dagger} \: ,
\nonumber\\
U \: c_{i\downarrow} \: U^{\dagger} \: & = & \: ( \: - \: 1 \: )^{N-1} \:
e^{i P_i} \: c_{i\downarrow}^{\dagger} \: ,
\nonumber\\
U \: c_{i\downarrow}^{\dagger} \: U^{\dagger} \: & = & \: ( \: - \: 1 \: )^{
N-1 } \: e^{ - i P_i } \: c_{i\downarrow} \: .
\label{uni1}
\end{eqnarray}
Based on Eq.(\ref{uni1}) the transformation of the particle number
operators can be given. As presented in Appendix 4., the following results are
obtained
\begin{eqnarray}
U \: n_{i\uparrow} \: U^{\dagger} \: & = & \: n_{i\uparrow} \: ,
\nonumber \\
U \: n_{i\downarrow} \: U^{\dagger} \: & = & \: 1 \: - \: n_{i\downarrow} \: .
\label{uni2}
\end{eqnarray}
Using Eqs.(\ref{uni1},\ref{uni2}) the complete Hamiltonian presented in
Eq.(\ref{eqH}) can be unitary transformed based on $U$.  For example in the
case $ \: t \: = \: X, \: $ $ P \: = \: R \: = \: 0$, and
$e^{i P_i } \: = \: 1$ for all $i$, we obtain the following final result
(see Appendix 5.)
\begin{eqnarray}
&& H' \: = \: - \: \tilde{t} \: \hat{t} \: + \: \tilde{U} \: \hat{U} \: + \:
\tilde{X} \: \hat{X} \: + \: \tilde{V} \: \hat{V} \: + \: \tilde{J_{z}} \:
\hat{J_{z}} \: + \: \frac{\tilde{J}_{xy}}{2} \: \hat{J}_{xy} \: + \:
\tilde{Y} \: \hat{Y}
\nonumber \\
&& \quad + \: \tilde{Q} \: \hat{Q} \: + \: \tilde{\mu} \: \sum_{j=1}^{N} \:
n_{j}\quad \: + \: \tilde{h} \: \sum_{j=1}^{N} \: ( \: n_{j\uparrow} \: - \:
n_{j\downarrow} \: ) \: + \: C \: .
\label{res1}
\end{eqnarray}
In this relation the renormalized coupling constants obtained after the
transformation are denoted with a tilde superscript their values being
\begin{eqnarray}
&& \tilde{t} \: = \: - \: t \: , \quad \quad \quad
\tilde{Y} \: = \: \frac{1}{2} \: J_{xy} \: ,
\nonumber\\
&& \tilde{X} \: = \: - \: X \: , \quad \quad
\tilde{Q} \: = \: Q \: ,
\nonumber\\
&& \tilde{U} \: = \: - \: U \: , \quad \quad
\: \tilde{\mu} \: = \: h \: ,
\nonumber\\
&& \tilde{V} \: = \: \frac{J_{z}}{4} \: , \quad \quad
\tilde{h} \: = \: \mu \: ,
\nonumber\\
&& \tilde{J}_{z} \: = \: 4 \: V \: , \quad \quad
\: C \: = \: ( \: \mu \: - \: h \: ) \: N \: ,
\nonumber\\
&& \tilde{J}_{xy} \: = \: 2 \: Y \: .
\label{res}
\end{eqnarray}

\section{The obtained results}

Based on the results presented in Chapter II. we are analysing
different concrete cases as follows.

\subsection{Ferromagnetic states}

De Boer et al (1995 a) in the study of the $\eta$-pairing ground state in
the extended Hubbard model used the following model Hamiltonian
\begin{eqnarray}
&& \hat{H_{1}} \: = \: - \: t \: \hat{t} \: + \: U \: \hat{U} \: + \: X \:
\hat{X} \: + \: V \: \hat{V} \: + \: J_{z} \: \hat{J}_{z} \: + \:
\frac{J_{xy}}{2} \: \hat{J}_{xy} \: + \: Y \: \hat{Y} \: + \: R \: \hat{R} \:
\nonumber\\
&& \quad + \: P \: \hat{P} \: + \: Q \: \hat{Q} \: + \: \mu \: \sum_{j=1}^{N}
\: n_{j} \: + \: h \: \sum_{j=1}^{N} \: ( \: n_{j\uparrow} \: - \:
n_{j\downarrow} \: ) \: ,
\end{eqnarray}
obtaining a superconducting $\eta$-pairing ground state at $ R \: = \: 0 , \:$
$ P \: = \: 0 ,$ in the following parameter space region
\begin{eqnarray}
&& V \: \leq \: 0 \: ,
\nonumber\\
&& 2 \: V \: = \: Y \: ,
\nonumber\\
&& t \: = \: X \: ,
\nonumber\\
&& - \: \frac{U}{z} \: \leq \: max \: [ \: V \: - \: \frac{J_{z}}{4} \: + \:
\frac{2 |h|}{z} \: ; \: V \: + \: \frac{J_{z}}{4} \: + \: \: | \:
\frac{J_{xy}}{2} \: | \; ;
\nonumber\\
&& 2 \: \frac{| \: h \: |}{z} \: + \: \frac{Q}{4} \: + \: 2 \: V \: + 2 \:
| \: t \: | \: ] \: ,
\end{eqnarray}
the ground state energy obtained being
\begin{eqnarray}
E_{N} \: = \: \frac{U \: N}{4} \: + \: \frac{Z \: N}{2} \: ( \: V \: - \:
\frac{P}{2} \: + \: \frac{Q}{16} \: ) \: + \: N_{e} \: ( \: 2 \: \mu \: + \:
\frac{P \: Z}{2} \: ) \: ,
\end{eqnarray}
where $N_e$ represents the number of electrons within the system.

Based on this result, taking into consideration that from Eq.(\ref{x1}) the
unitary transformed $\eta$-pairing state represents a ferromagnetic state,
using Eqs.(\ref{res1},\ref{res}) we obtain a ferromagnetic ground state in
the same model in the parameter space region $P \: = \: R \: = \: 0 $ and
\begin{eqnarray}
&& J_{z} \: \leq \: 0 \: ,
\nonumber\\
&& J_{z} \: = \: 2 \: J_{xy} \: ,
\nonumber\\
&& t \: = \: X \: ,
\nonumber\\
&& \frac{U}{z} \: \leq \: max \: [ \: \frac{J_{z}}{4} \: - \: V \: + \:
\frac{2}{z} \: | \: \mu \: | \: ; \: \frac{J_{z}}{4} \: + \: V \: + \:
| \: Y \: | \; ;
\nonumber\\
&& \frac{2}{z} \: | \: \mu \: | \: + \: \frac{Q}{4} \: + \: \frac{J_{z}}{2} \:
+ \: 2 \: | \: t \: | \: ] \: .
\end{eqnarray}
The ground state energy of this ferromagnetic state is given by
\begin{eqnarray}
\tilde{E_{N}} \: = \: - \: \frac{\tilde{U} \: N}{4} \: + \: \frac{Z \: N}{2}
\: ( \: \frac{\tilde{J_{z}}}{4} \: + \: \frac{\tilde{Q}}{16} \: ) \: + \: 2
\: N_{e} \: \tilde{h}
\end{eqnarray}
The position of the obtained ferromagnetic ground state in the $T \: = \: 0$
phase diagram of the model relative to the $\eta$-pairing state is illustrated
in Fig. 1.

The second Hamiltonian that we analyse concretely here has been studied by
Montorsi and Campbell (1996). The authors being interested in the
emergence possibilities of $\eta$-pairing superconductivity, used the
Hamiltonian
\begin{eqnarray}
&& H_{2} \: = \: - \: t \: \hat{t} \: + \: U \: \sum^{N}_{j=1} \:
n_{j\uparrow} \: n_{j\downarrow} \: + \: X \: \hat{X}
\nonumber\\
&& \quad + \: 2 \: \bar{V} \: \sum_{<j,k>} \: S_{j}^{'z} \: S_{k}^{'z} \: - \:
W \: \hat{W} \: - \: S \: \sum_{<j,k>} \: \vec{S}'_{j} \: \vec{S}'_{k}
\: + \: C \: .
\label{h2}
\end{eqnarray}
This Hamiltonian has been studied by Montorsi and Campbell (1996) obtaining
$\eta$-pairing superconducting ground state for $\bar{V} \: = \: 0$ in the
domain
\begin{eqnarray}
&& U \: \leq \: - \: A \: ,
\nonumber\\
&& S \: \geq \: 0 \: ,
\end{eqnarray}
where $A \: = \: Z \: ( \: 4 \: | \: X \: | \: + \: \frac{3}{2} \: |
\: W \: | \: )$, $\: Z$ representing the number of nearest neighbours.

In Appendix 6. we convert the Hubbard term from Eq.(\ref{h2}) into the form
it has in (\ref{U}). The unitary transformation presented in Chapter II
then gives a ferromagnetic ground state for $H_2$ in the region
\begin{eqnarray}
&& U \: \leq \: A'  \: ,
\label{tart}
\\
&& W \: \geq \: 0 \: ,
\end{eqnarray}
where $A' \: = \: Z \: ( \: 4 \: | \: X \: | \: + \: \frac{3}{2} \: | \:
S \: | \: )$, and $ \bar{V} \:  = \: 0, \: t \: = \: X$. The position of the
ferromagnetic domain in the phase diagram of the model is presented in
Fig. 2.

\subsection{Density wave states}

We are analysing now model Hamiltonians giving density wave states. We start
the presentation of our results with the Hamiltonian used by Strack and
Vollhardt (1993)
\begin{eqnarray}
H_{3} \: = \: - \: t \: \hat{t} \: + \: U \: \hat{U} \: + \: X \: \hat{X} \:
+ \: V \: \sum_{<j,i>,\sigma,\sigma'} \: n_{j\sigma} \: n_{i\sigma'} \: .
\label{xy5}
\end{eqnarray}
The last term from $H_3$ can be easily transformed into the form used in
Eq.(\ref{U}) (see  Appendix 7.). For the model from Eq.(\ref{xy5})
a paramagnetic insulator has been deduced in the pase diagram domain
\begin{eqnarray}
&& t \: = \: X \: ,
\nonumber\\
&& \frac{U}{Z} \: - \: 4 \: t \: - \: V \: \geq \: 0 \: ,
\label{xy1}
\end{eqnarray}
and a charge density wave region has been obtained in the region
(Strack and Vollhardt, 1993):
\begin{eqnarray}
&& t \: = \: X \: ,
\nonumber\\
&& \frac{U}{Z} \: + \: 4 \: t \: - \: V \: \leq \: 0 \: .
\label{xy2}
\end{eqnarray}
By the unitary transformation presented above we obtain instead of the
phase from Eq.(\ref{xy1}) an other paramagnetic insulator phase present in
the region
\begin{eqnarray}
&& t \: = \: X \: ,
\nonumber\\
&& - \: \frac{U}{Z} \: + \: 4 \: t \: - \: V^{+} \: \geq \: 0 \: ,
\nonumber\\
&& - \: \frac{U}{Z} \: + \: 4 \: t \: + \: V^{-} \: \geq \: 0 \: ,
\label{xy3}
\end{eqnarray}
and a spin density wave phase instead of the domain Eq.(\ref{xy2}) in the
region
\begin{eqnarray}
t \: = \: X \: ,
\nonumber\\
\frac{U}{Z} \: + \: 4 \: t \: + \: V^{+} \: \geq 0 \: ,
\nonumber\\
- \: \frac{U}{Z} \: - \: 4 \: t \: + \: V^{-} \: \leq \: 0 \: .
\label{xy4}
\end{eqnarray}
The results from Eqs.(\ref{xy3}, \ref{xy4}) are valid in a model similar to
Eq.(\ref{xy5}) in which however the density-density type coupling has been
replaced with
$V^{+} \: \hat{V}^{+} \: = \: V^{+} \: \sum_{<j,i>,\sigma} \: n_{j\sigma} \:
n_{i\sigma} \: ,$ and $ \: V^{-} \: \hat{V}^{-} \: = \: \sum_{<j,i>,\sigma}
\: n_{j\sigma} \: n_{i,-\sigma} \: $, see Appendix 7.
The results connected to the Hamiltonian $H_{3}$ are summarised in Fig. 3.

The next Hamiltonian that we analysed is connected to an extended Hubbard
model studied by de Boer et al. (1995 b)
\begin{eqnarray}
H_{4} \: = \: - \: t \: \hat{t} \: + \: U \: \hat{U} \: + \: X \: \hat{X}
\: + \: V \: \hat{V} \: + \: J_{z} \: \hat{J_{z}} \: + \: \frac{J_{xy}}{2}
\: \hat{J_{xy}} \: + \: Y \: \hat{Y} \: + \: \mu \: \sum_{j=1}^{N} \:
n_{j} \: .
\end{eqnarray}
Based on this Hamiltonian, de Boer et al. (1995 b.) proved the existence of
a SDW ground state in the following region of the parameter space
\begin{eqnarray}
&& J_{z} \: \geq \: 0 \: ,
\nonumber\\
&& J_{xy} \: = \: 0 \: ,
\nonumber\\
&& t \: = \: X \: ,
\nonumber\\
&& \frac{U}{z} \: \geq \: max ( \: \frac{J_{z}}{4} \: - \: V \: ; \: 2 \:
|t| \: - \: \frac{J_{z}}{2} \: ; \: V \: + \: |Y| \: - \: \frac{J_{z}}{4} \: )
\: .
\label{xxx1}
\end{eqnarray}
Using the presented unitary transformation, from Eq.(\ref{xxx1}) we obtain a
CDW ground state in the following region of the phase diagram of the same
model
\begin{eqnarray}
&& V \: \geq \: 0 \: ,
\nonumber\\
&& Y \: = \: 0 \: ,
\nonumber\\
&& t \: = \: X \: ,
\nonumber\\
&& - \: \frac{U}{z} \: \geq \: max ( \: \frac{J_{z}}{4} \: - \: \frac{V}{16}
\: ; \: 2 \: | \: t \: | \: - \: 2 \: | \: V \: | \: ; \: \frac{J_{z}}{2} \:
+ \: | \: \frac{J_{xy}}{2} \: - \: V \: ) \: .
\end{eqnarray}
de Boer and Schadschneider (1995 b.) obtained also a fully polarized
ferromagnetic state in the following region
\begin{eqnarray}
&& J_{z} \: \geq \: | \: J_{xy} \: | \: ,
\nonumber\\
&& \frac{U}{z} \: \geq \: max ( \: 2 \: | \: t \: | \: - \: \frac{J_{z}}{2}
\: ; \: \frac{J_{z}}{4} \: - \: V \: ; \: V \: + \: Y \: + \: \frac{J_{z}}{4}
\: ; \: V \: - \: Y \: + \: \frac{J_{z}}{4} \: ) \: .
\label{xxx2}
\end{eqnarray}
Interestingly, from the phase presented in Eq.(\ref{xxx2}) our unitary
transformation gives an insulating ground state in the domain
\begin{eqnarray}
&& \frac{V}{4} \: \geq \: | \: 2 \: Y \: | \: ,
\nonumber\\
&& - \: \frac{U}{z} \: \geq \: max ( \: 2 \: | \: t \: | \: + \: \frac{V}{8}
\: ; \: \frac{V}{16} \: - \: \frac{J_{z}}{4} \: ; \: \frac{J_{z}}{4} \: + \:
\frac{J_{xy}}{2} \: + \: \frac{V}{16} \: ) \: .
\end{eqnarray}
The obtained results related to the phase diagram of the Hamiltonian
$H_{4}$ are summarized in Fig. 4.

Another extended Hubbard Hamiltonian containing in it's ground state phase
diagram density waves has been analysed by Aligia et al. (1995)
\begin{eqnarray}
&& H_{5} \: = \: + \: U \: \sum^{N}_{j=1} \: n_{j\uparrow} \: n_{j\downarrow}
\: + \: V \: \sum_{<j,i>,\sigma\sigma'} \: n_{i\sigma} \: n_{j\sigma'} \: +
\nonumber \\
&& \sum_{<j,i>,\sigma} \: ( \: c^{\dagger}_{j\sigma} \: c_{i\sigma} \: + \:
c_{i\sigma}^{\dagger} \: c_{j\sigma} \: ) \: [ \: t_{AA} \: ( \: 1 \: - \:
n_{i,-\sigma} \: ) \: ( \: 1 \: - \: n_{j,-\sigma} \: ) \: + \: t_{BB} \:
n_{i,-\sigma} \: n_{j,-\sigma} \: ] \: .
\label{xxx3}
\end{eqnarray}
They found a region of parameters in the phase diagram where a CDW phase is
the ground state of the system in the domain
$V \: > \: V_{c} \: ,$ where
\begin{eqnarray}
V_{c} \: = \: 0 \: , \quad & if &  \quad U \: \leq \: - \: 2 \: Z \: t \: ,
\nonumber\\
V_{c} \: = \: \frac{U}{2Z} \: + \: t \: , \quad & if & \quad - \: 2 \: Z \:
t \: \leq \: U \: \leq \: 2 \: Z \: t \: ,
\nonumber\\
V_{c} \: = \: \frac{U}{Z} \: , \quad & if & \quad U \: \geq \: 2 \: Z \: t
\: .
\end{eqnarray}
Based on this result, using the presented unitary transformation (see
Appendix 8.) we have obtained a SDW ground state in the domain
$V^{+} \: > \: V_{c} \: , \quad V^{-} \: < \: V_{c} \: ,$ where
\begin{eqnarray}
V_{c} \: = \: 0 \: , \quad & if & \quad - \: U \: \leq \: - \: 2 \: Z \:
t \: ,
\nonumber\\
V_{c} \: = \: \frac{-U}{2Z} \: + \: t \: \quad & if & \quad - \: 2 \: Z
\: t \: \leq \: - \: U \: \leq \: 2 \: Z \: t \: ,
\nonumber\\
V_{c} \: = \: - \: \frac{U}{Z} \: \quad & if & \quad - \: U \: \geq \: 2
\: Z \: t \: .
\label{xxx4}
\end{eqnarray}
In Eq.(\ref{xxx4}) we have $ t \: = \: | \: t_{AA} \: | \: = \: | \: t_{BB}|$
The position of the deduced CDW phase in the phase diagram of the model is
presented in Fig. 5.

\section{Summary}
In this paper, we have presented a procedure that based on an explicitely
given unitary transformation allows the rigorous deduction of new phase
diagram domains starting from known ground state solutions of a given model.
The method is not model-dependent and can be applied for arbitrary dimensions.
We have applied this method to find new stable spin-density wave,
charge-density wave, ferromagnetic and a paramagnetic insulator domains of
different extended Hubbard type models.


\figure{Fig.1. Phase diagram regions in the coupling constant space
connected to Hamiltonian $H_{1}$. a) the superconducting state deduced
by De Boer et al. (1995 a.); b) the ferromagnetic ground state obtained in 
this paper.
\label{Fig.1.}}

\figure{Fig.2. Ground-state phase diagrams for the Hamiltonian $H_{2}$.
a) the superconducting state of Montorsi and Campbell (1996), b) the 
ferromagnetic ground state deduced in this paper.
\label{Fig.2.}}

\figure{Fig.3. Phase diagram regions in the coupling constant space connected
to Hamiltonian $H_{3}$. a) the ground states deduced by Strack and Vollhardt
(1993), densely dotted phase is CDW state, thinly dotted domain is 
paramagnetic insulator; b) ground states deduced in this paper: the densely 
dotted phase represents an SDW state and the thinly dotted region is an 
other paramagnetic insulator region.
\label{Fig.3.}}

\figure{Fig.4. Ground-state phase diagrams for the Hamiltonian $H_{4}$.
a) the SDW state obtained by de Boer et al. (1995 b.), b) the CDW phase 
obtained in this paper.
\label{Fig.4.}}

\figure{Fig.5. Phase diagram regions in the coupling constant space for the
Hamiltonian $H_{5}$. a) the CDW phase deduced by Aligia et al. (1995),
b) the SDW ground state deduced in this paper.
\label{Fig.5.}}

\newpage

\section{Appendices}

In this Chapter dedicated to Appendices we are presenting the mathematical
details of the calculation.

\subsection{Appendix 1.}

This appendix prove the unitary nature of the operator $U$. We must deduce
$U \: U^{\dagger} \: = \: 1$, where $U^{\dagger}$ is the adjoint of $U$.
Indeed, $U$
can be written as
\begin{eqnarray}
U \: = \: \prod_{i=1}^{N} \: z_{i} \: = \: z_{1} \: z_{2} \: ... \:
z_{N} \: ,
\end{eqnarray}
where $z^{+}_{i} \: = \: ( \: e^{-iPi} \: c_{i\downarrow} \: + \:
c_{i\downarrow}^{+} \: )$.
With this notation we have $ U^{\dagger} \: = \: z_{N}^{+} \: . \: . \: . \:
z_{1}^{+}$,
where
$z^{+}_{i} \: = \: ( \: e^{-iPi} \: c_{i\downarrow} \: + \:
c_{i\downarrow}^{+} \: )$. Then we find
$U \: U^{\dagger} \: = \: z_{1} \: . \: . \: . \: z_{N} \: z^{+}_{N} \: . \: .
\: . \: z_{1}^{+} $.
Since
\begin{eqnarray}
\: ( \: e^{iPi} \: c^{\dagger}_{i\downarrow} \: + \: c_{i\downarrow} \:
) \: ( \: e^{-iPi} \: c_{i\downarrow} \: +
\: c_{i\downarrow}^{\dagger}) \: = \: c_{i\downarrow}^{\dagger} \:
c_{i\downarrow} \: + \: c_{i\downarrow}
\: c_{i\downarrow}^{\dagger} \: = \: c_{i\downarrow}^{\dagger} \:
c_{i\downarrow} \: + \: 1 \: - \: c_{i\downarrow}^{\dagger}
c_{i\downarrow}=1  \:  ,
\end{eqnarray}
we obtain
$ z_{i} \: z_{i}^{+} \: = \: 1 $ for all  $i$. As a consequence
$U \: U^{\dagger} \: = \: 1 $, so $U$ is indeed an unitary transformation.

\subsection{Appendix 2}

In this appendix the effect of $U$ on the wave-functions is presented.
Let consider first the unitary transformation of the superconducting
($\eta$-pairing) wave-functions. If we consider a single doubly occupied
state present in the wave function at site $i$, the effect of the operator
$z_i$ on this state is to transform it in a single occuped state with spin up
as follows
\begin{eqnarray}
z_{i} \: | \: 0 \: , \: . \: . \: . \: , \: \uparrow\downarrow_{i} \: ,
\: 0 \: , \: . \: . \: . \: , \: 0 \: \rangle \: =
\: | \: 0 \: , \: . \: . \: . \: , \: \uparrow_{i} \: , \: 0 \: , \: .
\: . \: . \: , \: 0 \: \rangle \: .
\end{eqnarray}
The effect of $z_j$ on the same state for $i \ne j$ becomes
\begin{eqnarray}
z_{j} \: | \: 0 \: , \: . \: . \: . \: , \: 0_{j} \: , \: . \: . \: . \: , \:
\uparrow\downarrow_{i} \: , \: 0 \: , \: . \: . \: . \: , \: 0 \: \rangle \: =
 \: e^{iPj} \: | \: 0 \: , \: . \: . \: . \: , \: \downarrow_{j} \: , \: . \:
. \: . \: , \: \uparrow\downarrow_{i} \: , \: 0 \: , \: . \: . \: . \: , \: 0
\: \rangle \: .
\end{eqnarray}
From these equations it can be seen that the effect of $U$ on empty sites is
to introduce a sigle occupancy with spin down on these sites. However, in
case of doubly occupied sites containing an $\eta$-pair, the effect of $U$
is to transform these states in single occupied states with spin up. Based on
these relations, for a general wave function containing $N_{\eta}$ $\: \:
\eta$-pairs distributed on different sites we obtain
\begin{eqnarray}
&& U \: | \: 0 \: , \: . \: . \: . \: , \: 0 \: , \: \uparrow \:
\downarrow_{i1} \: , \: 0 \: , \: . \: . \: . \: , \: 0 \: ,
\: \uparrow\downarrow_{iN_{\eta}} \: , \: 0 \: , \: . \: . \: . \: , \: 0 \:
\rangle \: =
\nonumber\\
&& \frac{\prod_{j}^{N}e^{iPj}}{\prod_{i1}^{iN_{\eta}} \: e^{iPi}} \:
| \: \uparrow \: , \: \uparrow \: , \: . \: . \: . \: ,
\: \downarrow_{i1} \: , \: \uparrow \: , \: . \: . \: . \: ,
\: \downarrow_{iN_{\eta}} \: , \: . \: . \: . \: \uparrow \: \rangle \: .
\end{eqnarray}
Apart from a phase factor, this is a ferromagnetic state, if $N_{\eta} \:
{\neq} \: N/2$, where $N$ represents the number of lattice sites.

The study of density waves starts from the observation that $z_i$ transforms
differentially the single occupied states with up and down spin. Indeed,
for an up spin single occupied state we have
\begin{eqnarray}
z_{i} \: | \: . \: . \: . \: , \: \uparrow_{i} \: , \: . \: . \: . \: \rangle
\: = \: e^{iPi} \: | \: . \: . \: . \: , \: \uparrow\downarrow_{i} \: , \: .
\: . \: . \: \rangle \: ,
\end{eqnarray}
but for the spin down counterpart we obtain
\begin{eqnarray}
z_{i} \: | \: . \: . \: . \: , \: \downarrow_{i} \: , \: . \: . \: . \:
\rangle \: = \: | \: . \: . \: . \: , \: 0_{i} \: , \: . \: . \: . \:
\rangle \: .
\end{eqnarray}
Using this property, starting from a spin-density wave state, using $U$ we
find
\begin{eqnarray}
U \: | \: \downarrow \: , \: \uparrow \: , \: \downarrow \: , \: \uparrow \: ,
\: . \: . \: . \: \rangle \: =
 \: \prod_{i=1}^{N/2} \: e^{iP(2i)} \: | \: 0 \: ,
 \: \uparrow\downarrow \: , \: 0 \: , \: \uparrow\downarrow \: , \: . \: .
\: . \: \rangle \: .
\end{eqnarray}
The obtained state is a charge-density state.

\subsection{Appendix 3}

The transformation under $z_i$ of the creation and annihilation operators is
presented in this Appendix. Starting with annihilation operators, we have
for the spin up case
\begin{eqnarray}
z_{i} \: c_{i\uparrow} \: = \: ( \: e^{iPi} \: c_{i\downarrow}^{\dagger} \: +
\: c_{i\downarrow}) \: c_{i\uparrow} \: =
\: e^{iPi} \: c_{i\downarrow}^{\dagger} \: c_{i\uparrow} \: + \:
c_{i\downarrow} \: c_{i\uparrow} \: = \: - \: e^{iPi}
\: c_{i\uparrow} \: c_{i\downarrow}^{\dagger} \: - \: c_{i\uparrow} \:
c_{i\downarrow} \: ,
\end{eqnarray}
from where
\begin{eqnarray}
z_{i} \: c_{i\uparrow} \: z_{i}^{+} \: & = &
\: ( \: - \: e^{iPi} \: c_{i\uparrow} \: c_{i\downarrow}^{\dagger} \: - \:
c_{i\uparrow} \: c_{i\downarrow} \: )
\: ( \: e^{-iPi} \: c_{i\downarrow} \: + \: c_{i\downarrow}^{\dagger} \: ) \:
\nonumber\\
& = &
\: - \: c_{i\uparrow} \: c_{i\downarrow}^{\dagger} \: c_{i\downarrow} \: -
\: c_{i\uparrow} \: c_{i\downarrow} \: c_{i\downarrow}^{\dagger} \: = \: - \:
c_{i\uparrow}
\end{eqnarray}
arise. For different indices $i \: \ne \: j$ we find
$z_j \: c_{i,\uparrow} \: z_j^{\dagger} \: = \: - \: c_{i,\uparrow}$.
Similarly, for the opposite spin direction and different indices we obtain
$ z_{j} \: c_{i\downarrow} \: z_{j}^{+} \: = \: - \: c_{i\downarrow} $, and
for $i=j$ one have
\begin{eqnarray}
z_{i} \: c_{i\downarrow} \: = \: ( \: e^{iPi} \: c_{i\downarrow}^{\dagger} \:
+ \: c_{i\downarrow})  \: c_{i\downarrow} \: = \: e^{iPi} \:
c_{i\downarrow}^{\dagger} \: c_{i\downarrow} \: ,
\end{eqnarray}
\begin{eqnarray}
z_{i} \: c_{i\downarrow} \: z_{i}^{+} \: = \: e^{iPi} \:
c_{i\downarrow}^{\dagger} \: c_{i\downarrow}
\: ( \: e^{-iPi} \: c_{i\downarrow} \: + \: c_{i\downarrow}^{\dagger}) \: =
\: e^{iPi} \: c_{i\downarrow}^{\dagger} \: c_{i\downarrow}
\: c_{i\downarrow}^{\dagger} \: = \: e^{iPi} \: c_{i\downarrow}^{\dagger}
\end{eqnarray}
In the case of creation operators, for up spin, independent on indices we have
$z_{j} \: c_{i\uparrow}^{\dagger} \: z_{j}^{+} \: = \: - \:
c_{i\uparrow}^{\dagger}$ .
For down spin, in case of different indices we obtain
$z_{j} \: c_{i\downarrow}^{\dagger} \: z_{j}^{+} \: = \: - \:
c_{i\downarrow}^{\dagger} $ .
However, in case of $i \: = \: j $ the transformation gives
\begin{eqnarray}
z_{i} \: c_{i\downarrow}^{\dagger} \: = \: ( \: e^{iPi} \:
c_{i\downarrow}^{\dagger} \: + \: c_{i\downarrow} \: )
 \: c_{i\downarrow}^{\dagger} \: = \: c_{i\downarrow} \:
c_{i\downarrow}^{\dagger}\;\;,
\end{eqnarray}
\begin{eqnarray}
z_{i} \: c_{i\downarrow}^{\dagger} \: z_{i}^{+} \: = \: c_{i\downarrow} \:
c_{i\downarrow}^{\dagger}
(e^{-iPi}c_{i\downarrow}+c_{i\downarrow}^{\dagger}) =
\: e^{-iPi} \: c_{i\downarrow} \: c_{i\downarrow}^{\dagger} \:
c_{i\downarrow} \: =  \: e^{-iPi} \: c_{i\downarrow}
\end{eqnarray}
>From the presented results the transformation formulas from Eqs.(\ref{uni1})
arise.

\subsection{Appendix 4}

The unitary transformation generated by $U$ of the particle number operators
is presented below. We have
\begin{eqnarray}
U \: c_{i\uparrow}^{\dagger} \: c_{i\uparrow} \: U^{\dagger} \: =
\: U \: c_{i\uparrow}^{\dagger} \: U^{\dagger} \: U \: c_{i\uparrow} \:
U^{\dagger} \: =
\: ( \: - \: 1 \: )^{N} \: c_{i\uparrow}^{\dagger} \: ( \: - \: 1 \: )^{N} \:
c_{i\uparrow} \: = \: c_{i\uparrow}^{\dagger} \: c_{i\uparrow} \: =
\: n_{i\uparrow} \: ,
\end{eqnarray}
\begin{eqnarray}
&&U \: c_{i\downarrow}^{\dagger} \: c_{i\downarrow} \: U^{\dagger} \: =
\: U \: c_{i\downarrow}^{\dagger} \: U^{\dagger} \: U \: c_{i\downarrow} \:
U^{\dagger} \: =
\: ( \: - \: 1 \: )^{N-1} \: e^{-iPi} \: c_{i\downarrow}
\: ( \: - \: 1 \: )^{N-1} \: e^{iPi} \: c_{i\downarrow}^{\dagger} \: =
\nonumber \\
&&\quad\quad \: = \: c_{i\downarrow} \: c_{i\downarrow}^{\dagger} \: =
\: 1 \: - \: c_{i\downarrow}^{\dagger} \: c_{i\downarrow} \: = \: 1 \: - \:
n_{i\downarrow} \: .
\label{ntrans}
\end{eqnarray}
>From Eq.(\ref{ntrans}) $n_{i,\uparrow}$ remains unchanged, while
$n_{i,\downarrow}$
is transformed in $(1 \: - \: n_{i,\downarrow})$ under $U$.

\subsection{Appendix 5}

In this Appendix we are presenting the unitary transformation of different
terms from the Hamiltonian. We start the presentation with the Hubbard on-site
interaction
\begin{eqnarray}
&&U \: ( \: n_{j\uparrow} \: - \: 1/2 \: ) \: ( \: n_{j\downarrow} \: -
\: 1/2 \: ) \: U^{\dagger} \: = \: ( \: n_{j\uparrow} \: - \: 1/2 \: )
\: ( \: 1 \: - \: n_{j\downarrow} \: - \: 1/2 \: ) \: =
\nonumber\\
&&( \: n_{j\uparrow} \: - \: 1/2 \: ) \: ( \: 1/2 \: - \:
n_{j\downarrow} \: ) \: .
\label{Utransz}
\end{eqnarray}
It can be seen that the Hubbard interaction term changes the sign under the
unitary transformation used.

Concerning the hopping term, the following results are obtained
\begin{eqnarray}
&&U \: \hat{T}^{\uparrow}_{ij} \: U^{\dagger} \: = \: \hat{T}^{\uparrow}_{ij}
\nonumber \\
&&U \: \hat{T}^{\downarrow}_{ij} \: U^{\dagger} \: = \: - \: U \: ( \:
c^{\dagger}_{j\downarrow} \: c_{i\downarrow} \: +
\: c_{i\downarrow}^{\dagger} \: c_{j\downarrow}) \: U^{\dagger} \: =
\nonumber \\
&&- \: [ \: ( \: - \: 1 \: )^{N-1} \: e^{-iPj} \: ( \: - \: 1 \: )^{N-1} \:
e^{iPi} \: c_{j\downarrow}
\: c_{i\downarrow}^{\dagger} \: + \: ( \: - \: 1 \: )^{N-1} \: e^{-iPi} \: (
\: - \: 1 \: )^{N-1} \: e^{iPj}
\: c_{i\downarrow} \: c_{j\downarrow}^{\dagger} \: ] \:
\nonumber\\
&&= \: - \: e^{iP(j-i)} \: \hat{T} \: .
\label{Ttransz}
\end{eqnarray}
The transformation of the $\hat{X}$ operator yields
\begin{eqnarray}
&&U \: \hat{X}_{ij}^{\uparrow} \: U^{\dagger} \: =
U \: ( \: c_{j\uparrow}^{\dagger} \: c_{i\uparrow} \: + \:
c_{i\uparrow}^{\dagger} \: c_{j\uparrow} \: )
\: ( \: n_{j\downarrow} \: + \: n_{i\downarrow}) \: U^{\dagger} \: =
\nonumber \\
&&= \: ( \: c_{j\uparrow}^{\dagger} \: c_{i\uparrow} \: + \:
c_{i\uparrow}^{\dagger} \: c_{j\uparrow} \: ) \: ( \: 1 \: -
\: n_{j\downarrow} \: + \: 1 \: - \: n_{i\downarrow} \: ) \: =
\: - \: \hat{X}_{ij}^{\uparrow} \: - \: 2 \: \hat{T}_{ji}^{\uparrow} \: .
\nonumber \\
&&U \: \hat{X}_{ij}^{\downarrow} \: U^{\dagger} \: =
\: ( \: e^{i(Pj-Pi)} \: c_{j\downarrow} \: c^{\dagger}_{i\downarrow} \: +
\: e^{i(Pi-Pj)} \: c_{i\downarrow} \: c_{j\downarrow}^{\dagger} \: ) \: ( \:
n_{j\uparrow} \: + \: n_{i\uparrow} \: )
\nonumber\\
&&= \: -  \: e^{iP(j-i)} \: \hat{X}_{i,j}^{\downarrow} \: .
\label{Xtransz}
\end{eqnarray}
We consider now the $\hat{J}_{z}$ and $\hat{V}$ operatorial terms
\begin{eqnarray}
&&U \: \hat{J}_{z} \: U^{\dagger} \: = \: \frac{1}{4} \: U \: ( \:
n_{j\uparrow} \: - \: n_{j\downarrow} \: ) \: ( \: n_{i\uparrow}
\: - \: n_{i\downarrow}) \: U^{\dagger}
\nonumber\\
&&= \: \frac{1}{4} \: ( \:
n_{j\uparrow} \: - \: 1 \: + \: n_{j\downarrow} \: )
\: ( \: n_{i\uparrow} \: - \: 1 \: + \: n_{i\downarrow} \: ) \: =
\nonumber \\
&&\frac{1}{4} \: ( \: n_{j} \: - \: 1 \: ) \: ( \: n_{i} \: -
\: 1 \: ) \: = \: \frac{1}{4} \: \hat{V} \: .
\nonumber \\
&&U \: \hat{V} \: U^{\dagger} \: = \: U \: ( \: n_{j} \: - \: 1 \: ) \: ( \:
n_{i} \: - \: 1 \: ) \: U^{\dagger} \: = \: U \: ( \: n_{j\uparrow} \: + \:
n_{j\downarrow} \: - \: 1 \: )
\: ( \: n_{i\uparrow} \: + \: n_{i\downarrow} \: - \: 1 \: ) \: U^{\dagger} \:
\nonumber \\
&&= \: ( \: n_{j\uparrow} \: - \: n_{j\downarrow} \: ) \: ( \: n_{i\uparrow}
\: - \: n_{i\downarrow}) \: = \: 4 \: \hat{J}_{z}
\label{J-V}
\end{eqnarray}
>From Eq.(\ref{J-V}) the $\hat{J}_{z}$ and $\hat{V}$ operators transform into
each
other.

The following terms analysed are $\hat{J}_{xy}$ and $\hat{Y}$. We obtain
\begin{eqnarray}
&&U \: \hat{J}_{xy\;ij} \: U^{\dagger}= \: U \: \frac{1}{2} \: ( \: ( \:
S_{j}^{+} \: S_{i}^{-} \: +
 \: S_{j}^{-} \: S_{i}^{+}) \: U^{\dagger} \: =
\nonumber\\
&&U \: \frac{1}{2} \: ( \: c_{j\uparrow}^{\dagger} \: c_{j\downarrow} \:
c_{i\downarrow}^{\dagger} \: c_{i\uparrow} \: +
\: c_{j\downarrow}^{\dagger} \: c_{j\uparrow} \: c_{i\uparrow}^{\dagger} \:
c_{i\downarrow}) \: U^{\dagger} \: =
\nonumber\\
&&\frac{1}{2} \: ( \: c_{j\uparrow}^{\dagger} \: e^{iPj} \:
c_{j\downarrow}^{\dagger} \: e^{-iPi} \: c_{i\downarrow}
\: c_{i\uparrow} \: + \: e^{-iPj} \: c_{j\downarrow} \: c_{j\uparrow} \:
c_{i\uparrow}^{\dagger} \: e^{iPi}
\: c_{i\downarrow}^{\dagger} \: ) \: =
\nonumber\\
&&\frac{1}{2} \: ( \: e^{i(Pj-Pi)} \: c_{j\uparrow}^{\dagger} \:
c_{j\downarrow}^{\dagger}  \: c_{i\downarrow} \: c_{i\uparrow} \: +
\: e^{i(Pi-Pj)} \: c_{i\uparrow}^{\dagger} \: c_{i\downarrow}^{\dagger} \:
c_{j\downarrow} \: c_{j\uparrow}) \: =
\frac{1}{2} \: e^{i(Pj-Pi)} \: \hat{Y} \: .
\nonumber\\
&&U \: \hat{Y}_{ij} \: U^{\dagger} \: = \: U \: ( \: c_{j\uparrow}^{\dagger}
\: c_{j\downarrow}^{\dagger} \: c_{i\downarrow} \: c_{i\uparrow} \: +
\: c_{i\uparrow}^{\dagger} \: c_{i\downarrow}^{\dagger} \: c_{j\downarrow} \:
c_{j\uparrow}) \: U^{\dagger} \: =
\nonumber\\
&&e^{i(Pi-Pj)} \: c_{j\uparrow}^{\dagger} \: c_{j\downarrow} \:
c_{i\downarrow}^{\dagger} \: c_{i\uparrow} \: +
e^{i(Pj-Pi)} \: c_{j\downarrow}^{\dagger} \: c_{j\uparrow} \:
c_{i\uparrow}^{\dagger} \: c_{i\downarrow} \: =
2 \: e^{i(Pj-Pi)} \: \hat{J}_{xy\;ij} \: .
\label{J-Y}
\end{eqnarray}
>From Eq.(\ref{J-Y}), the $\hat{J}_{xy}$ and $\hat{Y}$ operators transform into
each other if the phase factor $Pi \: = \: 0$ is considered. Because of this
result and Eq.(\ref{J-V}), the $\hat{W}$ and $\hat{S}$ operators transform
into each other as well. Indeed,
\begin{eqnarray}
&&U \: S_{j}^{'z} \: S_{i}^{'z} \: U^{\dagger} \: = \: S_{j}^{z} \:
S_{i}^{z} \: ,
\nonumber\\
&&U \: S_{j}^{'+} \: S_{i}^{'-} \: U^{\dagger} \: = \: e^{i(Pj-Pi)} \:
S_{j}^{+} \: S_{i}^{-} \: ,
\nonumber\\
&&U \: S_{j}^{'-} \: S_{i}^{'+} \: U^{\dagger} \: = \: e^{i(Pj-Pi)} \:
S_{j}^{-} \: S_{i}^{+} \: ,
\end{eqnarray}
and
\begin{eqnarray}
&&U \: S_{j}^{z} \: S_{i}^{z} \: U^{\dagger} \: = \: S_{j}^{'z} \:
S_{i}^{'z} \: ,
\nonumber\\
&&U \: S_{j}^{+} \: S_{i}^{-} \: U^{\dagger} \: = \: e^{i(Pj-Pi)}  \:
S_{j}^{'+} \: S_{i}^{'-} \: ,
\nonumber\\
&&U \: S_{j}^{-} \: S_{i}^{+} \: U^{\dagger} \: = \: e^{iP(i-j)} \:
S_{j}^{'-} \: S_{i}^{'+} \: .
\end{eqnarray}
We consider now the transformation of the $\hat{R}$ operator
\begin{eqnarray}
&&U \: \hat{R}_{ji}^{\uparrow}U^{\dagger} \: = \: U \: ( \:
c_{j\uparrow}^{\dagger} \: c_{i\uparrow} \: + \: c_{i\uparrow}^{\dagger}
\: c_{j\uparrow}) \: n_{j\downarrow} \: n_{i\downarrow} \: U^{\dagger} \: =
\nonumber\\
&& ( \: c_{j\uparrow}^{\dagger} \: c_{i\uparrow} \: +
\: c_{i\uparrow}^{\dagger} \: c_{j\uparrow} \: ) \: ( \: 1 \: - \:
n_{j\downarrow} \: ) \: ( \: 1 \: - \: n_{i\downarrow}) \: =
\nonumber\\
&&\hat{R}_{ji}^{\uparrow} \: - \: ( \: c_{j\uparrow}^{\dagger} \:
c_{i\uparrow} \: +  \: c_{i\uparrow}^{\dagger} \: c_{j\uparrow} \: ) \: -
\: ( \: c_{j\uparrow}^{\dagger} \: c_{i\uparrow} \: + \:
c_{i\uparrow}^{\dagger} \: c_{j\uparrow} \: )
\: ( \: n_{j\downarrow} \: + \: n_{i\downarrow}) \: =
\nonumber\\
&&\hat{R}_{ji}^{\uparrow} \: - \: \hat{T}_{ji}^{\uparrow} \: -
\: \hat{X}_{ji}^{\uparrow} \: .
\nonumber\\
&&U \: \hat{R}_{ji}^{\downarrow} \: U^{\dagger} \: =
U \: ( \: c_{j\downarrow}^{\dagger} \: c_{i\downarrow} \: + \:
c_{i\downarrow}^{\dagger} \: c_{j\downarrow} \: )
\: n_{j\uparrow} \: n_{i\uparrow} \: U^{\dagger} \: =
\nonumber\\
&&( \: e^{i(Pi-Pj)} \: c_{j\downarrow} \: c_{i\downarrow}^{\dagger} \: +
\: e^{i(Pi-Pj)} \: c_{i\downarrow} \: c_{j\downarrow}^{\dagger}) \:
n_{j\uparrow} \: n_{i\uparrow} \: =
\nonumber\\
&&e^{i(Pi-Pj)} \: \hat{R}_{ji}^{\downarrow} \: .
\label{ezR}
\end{eqnarray}
We consider now the operator $\hat{P}$. This operator is build up from a sum
of $\hat{P}_{ij}$ terms whose expression is given by
\begin{eqnarray}
&&\hat{P}_{ij} \: = \: ( \: n_{j\uparrow} \: - \: \frac{1}{2} \: ) \: ( \:
n_{j\downarrow} \: -
 \: \frac{1}{2} \: ) \: ( \: n_{i\uparrow} \: + \: n_{i\downarrow}- \:
\frac{1}{2} \: - \: \frac{1}{2} \: ) \: +
\nonumber\\
&&( \: n_{i\uparrow} \: - \: \frac{1}{2} \: ) \: ( \: n_{i\downarrow} \: -
\: \frac{1}{2} \: ) \: ( \: n_{j\uparrow} \: - \: \frac{1}{2} \: +
\: n_{j\downarrow}- \: \frac{1}{2} \: ) \: .
\end{eqnarray}
This operator can be decomposed in  the following components
\begin{eqnarray}
&&\hat{P}_{ij}^{+} \: =
\: ( \: n_{j\uparrow} \: - \: \frac{1}{2} \: ) \: ( \: n_{j\downarrow} \: - \:
\frac{1}{2} \: ) \:
( \: n_{i\downarrow}- \: \frac{1}{2} \: ) \: + \: ( \: n_{i\uparrow} \: - \:
\frac{1}{2} \: ) \: ( \: n_{i\downarrow} \: - \: \frac{1}{2} \: ) \: ( \:
n_{j\downarrow} - \: \frac{1}{2} \: ) \: ,
\nonumber\\
&&\hat{P}_{ij}^{-} \: = \: ( \: n_{j\uparrow} \: - \: \frac{1}{2} \: ) \: ( \:
n_{j\downarrow} \: - \: \frac{1}{2} \: ) \: ( \: n_{i\uparrow}- \: \frac{1}{2}
\: ) \:
+ \: ( \: n_{i\uparrow} \: - \: \frac{1}{2} \: ) \: ( \: n_{i\downarrow} \: -
\: \frac{1}{2} \: ) \: ( \: n_{j\uparrow}- \: \frac{1}{2} \: ) \: .
\end{eqnarray}
The unitary transformation of $\hat{P}^{+}$ and $\hat{P}^{-}$ operators gives
the following results
\begin{eqnarray}
&&U \: \hat{P}_{ij}^{+} \: U^{\dagger} \: = \: ( \: n_{j\uparrow} \: - \:
\frac{1}{2} \: ) \: ( \: - \: n_{j\downarrow} \: + \: \frac{1}{2} \: ) \: ( \:
- \: n_{i\downarrow} \: + \: \frac{1}{2} \: ) \: +
\nonumber\\
&& \:  \:  \:  \:  \:  \:  \:  \:  \: + \: ( \: n_{i\uparrow} \: - \:
\frac{1}{2} \: ) \: ( \: - \: n_{i\downarrow} \: - \: \frac{1}{2} \: ) \:
( \: - \: n_{j\downarrow}+ \: \frac{1}{2} \: ) \: = \: \hat{P}_{ij}^{+} \: .
\nonumber\\
&&U \: \hat{P}_{ij}^{-} \: U^{\dagger} \: = \: ( \: n_{j\uparrow} \: - \:
\frac{1}{2} \: ) \: ( \: - \: n_{j\downarrow} \: + \: \frac{1}{2} \: ) \:
( \: n_{i\uparrow}- \: \frac{1}{2} \: ) \: +
\nonumber\\
&& \:  \:  \:  \:  \:  \:  \:  \:  \: + \: ( \: n_{i\uparrow} \: - \:
\frac{1}{2} \: ) \: ( \: - \: n_{i\downarrow} \: + \: \frac{1}{2} \: ) \: ( \:
n_{j\uparrow} \: - \: \frac{1}{2} \: ) \: = \: -\hat{P}_{ij}^{-} \: .
\end{eqnarray}
Based on these results one can see, that $\hat{P}_{ij}^{+}$ remains unaltered,
while $\hat{P}_{ij}^{-}$ changes the sign under the unitary transformation
used.

The following operatorial term transformed is $\hat{Q}$. We have
\begin{eqnarray}
&&U \: \hat{Q}_{ij} \: U^{\dagger} \: = \: U \: ( \: n_{j\uparrow} \: - \:
\frac{1}{2} \: ) \: ( \: n_{j\downarrow} \: - \: \frac{1}{2} \: )
\: ( \: n_{i\uparrow} \: - \: \frac{1}{2} \: ) \: ( \: n_{i\downarrow} \:
- \: \frac{1}{2}) \: U^{\dagger} \: = \:
\nonumber \\
&&\quad\quad
( \: n_{j\uparrow} \: - \: \frac{1}{2} \: ) \: ( \: - \: n_{j\downarrow} \: +
\: \frac{1}{2}) \: ( \: n_{i\uparrow} \: - \: \frac{1}{2})
 \: ( \: - \: n_{i\downarrow} \: + \: \frac{1}{2}) \: = \: \hat{Q}_{ij} \: .
\end{eqnarray}
As a consequence, the operator $\hat{Q}$ transform into itself under the
unitary transformation generated by $U$.

We consider now the chemical potential $\mu$ and external field $h$.
Introducing the notations  $\mu_{j} \: = \: n_{j\uparrow} \: +
\: n_{j\downarrow}$ and $h_{j} \: = \: n_{j\uparrow} \: - \: n_{j\downarrow}$,
we obtain
\begin{eqnarray}
&&U \: \mu_{j}U^{\dagger} \: = \: n_{j\uparrow} \: + \: 1 \: - \:
n_{j\downarrow} \: = \: h_{j} \: + \: 1 \: ,
\nonumber\\
&&U \: h_{j} \: U^{\dagger} \: = \: n_{j\uparrow} \: - \: 1 \: + \:
n_{j\downarrow} \: = \: \mu_{j} \: - \: 1 \: .
\end{eqnarray}
We mention that after transforming all terms of the Hamiltonian, driven by
the aim to remain in the same class of model, we consider the transformed
Hamiltonian identical with the original. This will be realized if for all
lattice sites we have $e^{iPi} \: = \: 1$, and $\tilde{X^{\uparrow}} \: =
\: \tilde{X^{\downarrow}}$, $ \: \tilde{t^{\downarrow}} \: = \:
\tilde{t^{\uparrow}}$,  $ \: \tilde{P}^{+} \: = \: \tilde{P}^{-}$ . Since
$\tilde{X^{\uparrow}} \: = \: - \: X \: - \: R$ and $\tilde{X^{\downarrow}} \:
= \: - \: X$, from Eq.(\ref{ezR}) we get the condition $R \: = \: 0$ .
Besides, as $\tilde{t^{\uparrow}} \: = \: t \: - \: 2 \: X$ and
$\tilde{t^{\downarrow}} \: = \: - \: t$ we get $t \: = \: X$ .
We also mention that during this paper $Pi \: = \: 0$ has been used.

\subsection{Appendix 6}

In this Appendix we are presenting the relation that connects the form of the
Hubbard interaction analysed in Appendix 4. and the standard form of the
on-site interaction expressed via a $\sum_i n_{i,\uparrow} n_{i,\downarrow}$
product. We have
\begin{eqnarray}
&&\sum_{i} \: ( \: n_{i\uparrow} \: - \: \frac{1}{2} \: ) \: ( \:
n_{i\downarrow} \: - \: \frac{1}{2}) \: =
\: \sum_{i} \: ( \: n_{i\uparrow} \: n_{i\downarrow} \: - \: \frac{1}{2}
\: ( \: n_{i\uparrow} \: +
\: n_{i\downarrow}) \: + \frac{1}{4} \: )
\nonumber\\
&&= \: \sum_{i} \: n_{i\uparrow} \: n_{i\downarrow} \: - \: \frac{1}{2}
\sum_{i\sigma}n_{i\sigma} \: + \: \frac{1}{4} \: \sum_{i} \: 1
\: = \: \sum_{i} \: n_{i\uparrow} \: n_{i\downarrow} \: - \: \frac{1}{2} \:
N_{e} \: + \: \frac{1}{4} \: N \: ,
\end{eqnarray}
where $N_{e}$ represents the number of electrons within the system and $N$
is the number of lattice sites. Denoting $N/4 \: - \: N_e/2 \: = \: C$, we
obtain
\begin{eqnarray}
U \: \sum^{N}_{j=1} \: n_{j\uparrow} \: n_{j\downarrow} \: = \: U \: \sum_{i}
\: ( \: n_{i\uparrow} \: - \: \frac{1}{2}) \: ( \: n_{i\downarrow} \: -
\: \frac{1}{2} \: ) \: + \: U \: C \: .
\label{Ual}
\end{eqnarray}
The result obtained in Eq.(\ref{Ual}) is used for the study of the
Hamiltonian $H_2$ presented in Eq.(\ref{h2}). Using Eq.(\ref{Ual}), the
Hamiltonian $H_2$ becomes
\begin{eqnarray}
&&H' \: = \: - \: t \: \hat{t} \: + \: U \: \sum^{N}_{j=1} \: ( \:
n_{j\uparrow} \: - \: \frac{1}{2} \: ) \: ( \: n_{j\downarrow} \: -
\: \frac{1}{2}) \: + \: X \: \hat{X}
\nonumber\\
&&+ \: 2 \: V \: \sum_{<j,k>} \: S_{j}^{'z} \: S_{k}^{'z} \: - \: W \:
\hat{W} \: - \: S \: \sum_{<j,k>} \: \vec{S}'_{j} \: \vec{S}'_{k} \:
+ \: C' \: .
\end{eqnarray}

\subsection{Appendix 7}

This Appendix presents in detail the transformation of the Hamiltonian $H_3$
from Eq.(\ref{xy5}) in a form more convenient for our study. We have
converted the last term of $H_{3}$ with the help of the following identity:
\begin{eqnarray}
&&\sum_{<j,i>,\sigma,\sigma'} \: ( \: n_{j\sigma} \: - \: \frac{1}{2} \: )
\: ( \: n_{i\sigma'} \: - \: \frac{1}{2} \: ) \: =
\: \sum_{<j,i>,\sigma} \: ( \: ( \: n_{j\sigma} \: n_{i,-\sigma} \: - \:
\frac{1}{2} \: n_{i\sigma} \: -
\: \frac{1}{2} \: n_{j,-\sigma} \: + \: \frac{1}{4}) \: +
\nonumber\\
&& ( \: n_{j\sigma} \: n_{i\sigma} \: - \: \frac{1}{2} \: n_{i\sigma} \: - \:
\frac{1}{2} \: n_{j\sigma} \: + \: \frac{1}{4} \: ) \: ) \: =
\nonumber\\
&& \: \sum_{<j,i>,\sigma,\sigma'} \: n_{i\sigma'} \: n_{j\sigma}
\: - \: \frac{1}{2} \: \sum_{<j,i>,\sigma} \: ( \: n_{i\sigma} \: + \:
n_{j,-\sigma} \: + \: n_{i\sigma}
\: + \: n_{j\sigma}) \: + \: 4 \: \frac{1}{4} \: =
\nonumber\\
&&\sum_{<j,i>,\sigma,\sigma'} \: n_{i\sigma'} \: n_{j\sigma} \: -
\: \sum_{<j,i>,\sigma} \: n_{i\sigma} \: -
\: \frac{1}{2} \: \sum_{<j,i>} \: ( \: n_{j\uparrow} \: + \: n_{j\downarrow}
\: + \: n_{j\uparrow} \: + \: n_{j\downarrow}) \: + \: 1 \: =
\nonumber\\
&&\sum_{<j,i>,\sigma,\sigma'} \: n_{i\sigma'} \: n_{j\sigma} \: - \: 2 \:
N_{e} \: + \: 1 \: .
\label{ez1}
\end{eqnarray}
Based on Eq.(\ref{ez1}) we have
\begin{eqnarray}
\sum_{<j,i>,\sigma,\sigma'} \: n_{i\sigma'} \: n_{j\sigma} \: =
\: \sum_{<j,i>,\sigma,\sigma'} \: ( \: n_{j\sigma} \: - \: \frac{1}{2} \: ) \:
( \: n_{i\sigma'} \: - \: \frac{1}{2} \: ) \: + \: const.
\label{Val}
\end{eqnarray}
Using Eq.(\ref{Val}), the Hamiltonian $H_3$ becomes
\begin{eqnarray}
H \: = \: - \: t \: \hat{t} \: +\;U \: \hat{U} \: + \: X \: \hat{X} \:
+ \: V \: \hat{V} \: + \: const.
\label{Val1}
\end{eqnarray}
Using the form presented in Eq.(\ref{Val1}) different terms of the
Hamiltonian can be unitary transformed. The transformation of the operatorial
terms presented in Eq.(\ref{Val}) gives
\begin{eqnarray}
U \: ( \: n_{j\uparrow} \: - \: \frac{1}{2} \: ) \: ( \: n_{i\uparrow} \: - \:
\frac{1}{2}) \: U^{\dagger} \: =
\: ( \: n_{j\uparrow} \: - \: \frac{1}{2} \: ) \: ( \: n_{i\uparrow} \: - \:
\frac{1}{2} \: ) \: ,
\nonumber\\
U \: ( \: n_{j\uparrow} \: - \: \frac{1}{2} \: ) \: ( \: n_{i\downarrow} \: -
\: \frac{1}{2}) \: U^{\dagger} \: =
\: ( \: n_{j\uparrow} \: - \: \frac{1}{2} \: ) \: ( \: \frac{1}{2} \: - \:
n_{i\downarrow} \: ) \: ,
\nonumber\\
U \: ( \: n_{j\downarrow} \: - \: \frac{1}{2} \: ) \: ( \: n_{i\uparrow} \: -
\: \frac{1}{2}) \: U^{\dagger} \: =
\: ( \: \frac{1}{2} \: - \: n_{j\downarrow} \: ) \: ( \: n_{i\uparrow} \: - \:
\frac{1}{2} \: ) \: ,
\nonumber\\
U \: ( \: n_{j\downarrow} \: - \: \frac{1}{2} \: ) \: ( \: n_{i\downarrow}
\: - \: \frac{1}{2}) \: U^{\dagger} \: =
\: ( \: \frac{1}{2} \: - \: n_{j\downarrow} \: ) \: ( \: \frac{1}{2} \: - \:
n_{i\downarrow}) \: .
\end{eqnarray}
Decomposing the contribution presented in Eq.(\ref{Val}) is spin dependent
terms, new notation has to be introduced
\begin{eqnarray}
V^{+} \: = \: \sum_{<j,i>,\sigma} \: n_{j\sigma} \: n_{i\sigma} \: , \quad
V^{-} \: = \: \sum_{<j,i>,\sigma} \: n_{j\sigma} \: n_{i,-\sigma}
\label{ez2}
\end{eqnarray}
The unitary transformation of the operatorial components from Eq.(\ref{ez2})
gives
$U \: V^{+} \: U^{\dagger} \: = \: V^{+}$, and
$U \: V^{-} \: U^{\dagger} \: = \: - \: V^{-}$.

\subsection{Appendix 8}
The unitary transformation of the Hamiltonian $H_5$ from Eq.(\ref{xxx3}) is
presented in this Appendix. Before starting the unitary transformation we
mention that the first two terms of $H_5$ may be rewritten based on
Eqs.(\ref{Ual}, \ref{Val}). For the third term of $H_5$ we are using the
expression
\begin{eqnarray}
&& \: \sum_{<i,j>} \: ( \: c^{\dagger}_{j\uparrow} \: c_{i\uparrow} \: +
 \: c_{i\uparrow}^{\dagger} \: c_{j\uparrow}) \: [ \: t_{AA} \: n_{i\downarrow}
\: n_{j\downarrow} \:
+ \: t_{BB} \: ( \: 1 \: - \: n_{i\downarrow} \: ) \: ( \: 1 \: - \:
n_{j\downarrow} \: ) \: ] \: +
\nonumber\\
&& \: \sum_{<i,j>} \: ( \: c^{\dagger}_{j\downarrow} \: c_{i\downarrow} \: +
 \: c_{i\downarrow}^{\dagger} \: c_{j\downarrow} \: ) \: [ \: ( \: ( \: - \: 1
\: ) \: t_{AA} \: ( \: 1 \: - \: n_{i\uparrow} \: ) \: ( \: 1 \: - \:
n_{j\uparrow} \: ) \: + \:
\nonumber\\
&& ( \: - \: 1 \: ) \: t_{BB} \: n_{i\uparrow} \: n_{j\uparrow}] \: .
\end{eqnarray}
The unitary transformation of the new operatorial terms not encountered up to
this moment are
\begin{eqnarray}
U \: ( \: c^{\dagger}_{j\uparrow} \: c_{i\uparrow} \: +
 \: c_{i\uparrow}^{\dagger} \: c_{j\uparrow}) \: U^{\dagger} \: &=&
 \: ( \: c^{\dagger}_{j\uparrow} \: c_{i\uparrow} \: +
 \: c_{i\uparrow}^{\dagger} \: c_{j\uparrow} \: )
 \nonumber \\
U \: ( \: 1 \: - \: n_{i\downarrow} \: ) \: ( \: 1 \: - \: n_{j\downarrow}) \:
U^{\dagger} \: &=& \: n_{i\downarrow} \: n_{j\downarrow}
 \nonumber \\
U \: n_{i\downarrow} \: n_{j\downarrow} \: U^{\dagger} \: &=& \: ( \: 1 \: - \:
n_{i\downarrow} \: ) \: ( \: 1 \: - \: n_{j\downarrow} \: )
 \nonumber \\
U \: ( \: c^{\dagger}_{j\downarrow} \: c_{i\downarrow} \: +
 \: c_{i\downarrow}^{\dagger} \: c_{j\downarrow} \: ) \: U^{\dagger} \: &=&
 \: - \: ( \: c^{\dagger}_{j\downarrow} \: c_{i\downarrow} \: + \:
c_{i\downarrow}^{\dagger} \: c_{j\downarrow})
\nonumber \\
U \: ( \: 1 \: - \: n_{i\uparrow} \: ) \: ( \: 1 \: - \: n_{j\uparrow}) \:
U^{\dagger}& \: =& \: ( \: 1 \: - \: n_{i\uparrow} \: ) \: ( \: 1 \: - \:
n_{j\uparrow} \: )
\nonumber \\
U \: n_{i\uparrow} \: n_{j\uparrow} \: U^{\dagger} \: &=& \: n_{i\uparrow} \:
n_{j\uparrow}
\end{eqnarray}
We mention that during the paper we assumed $t_{AA} \: = \: - \: t_{BB}$,
and the following notation has been used
\begin{eqnarray}
t \: = \: | \: t_{AA} \: | \: = \: | \: t_{BB}| \: = \: | \: \tilde{t}_{AA}
\: | \: = \: | \: \tilde{t}_{BB} \: | \: .
\end{eqnarray}

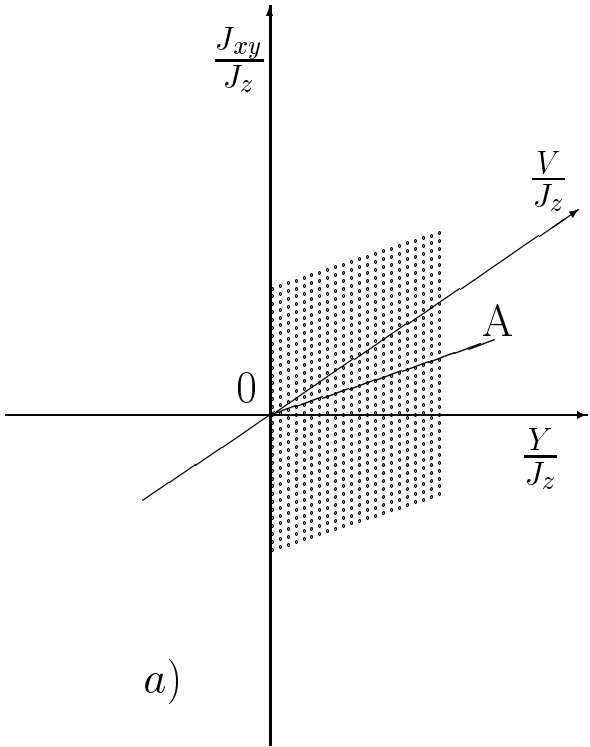
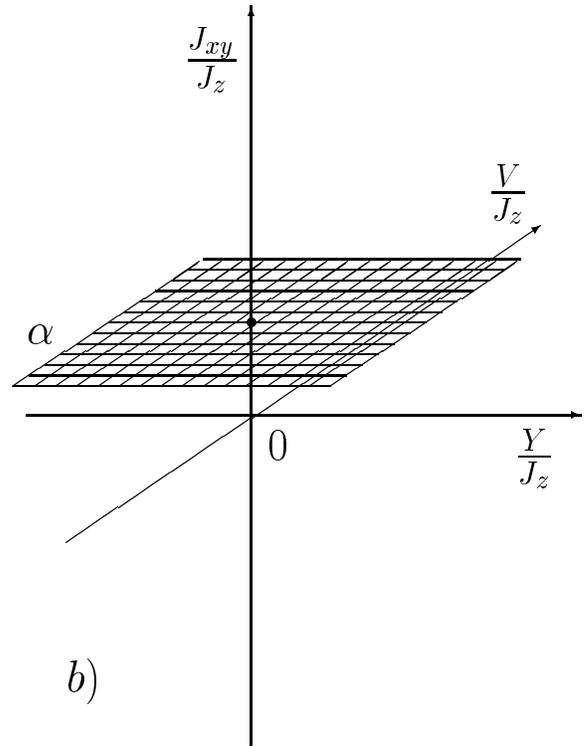
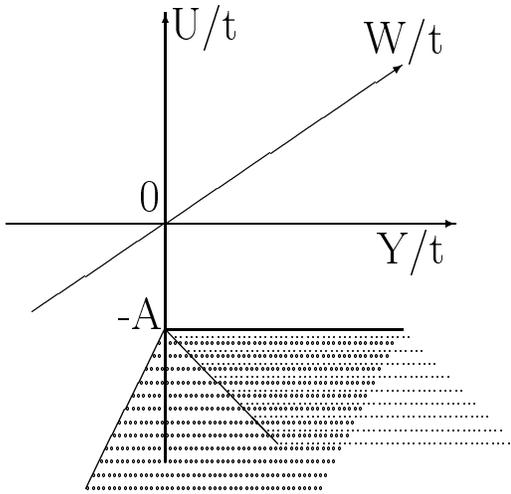
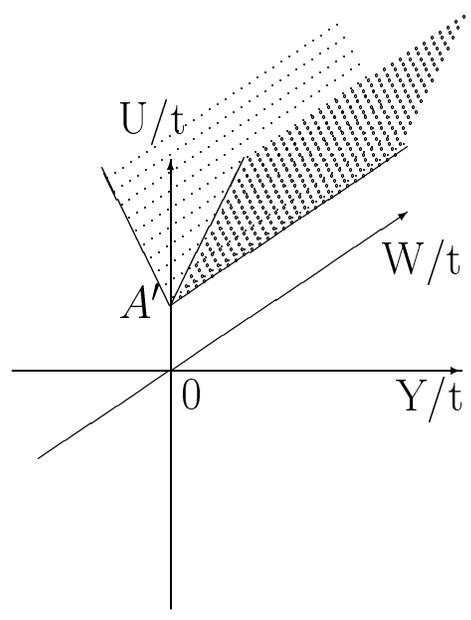
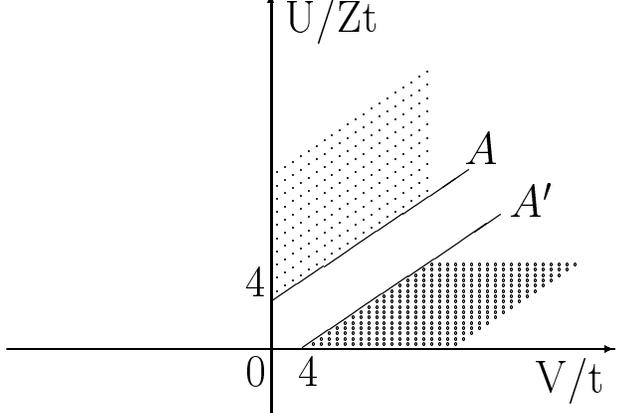
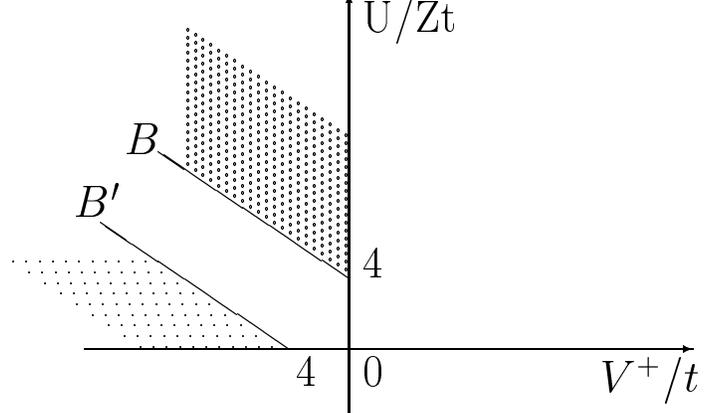
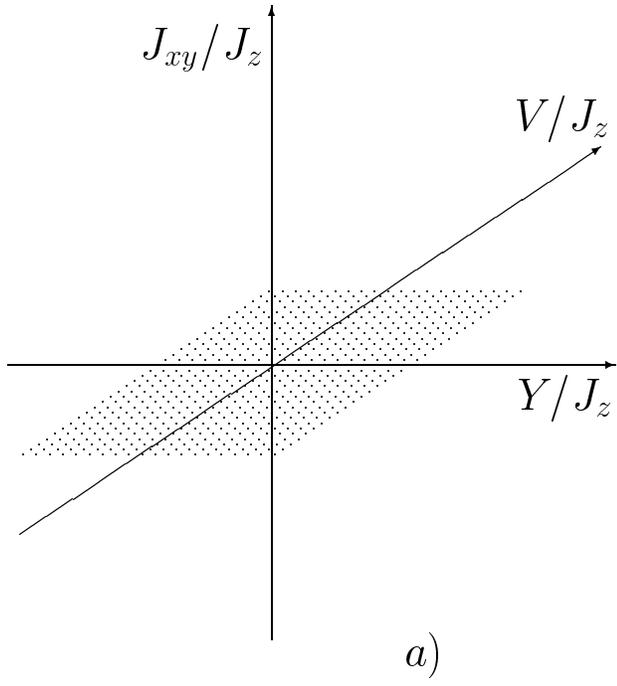
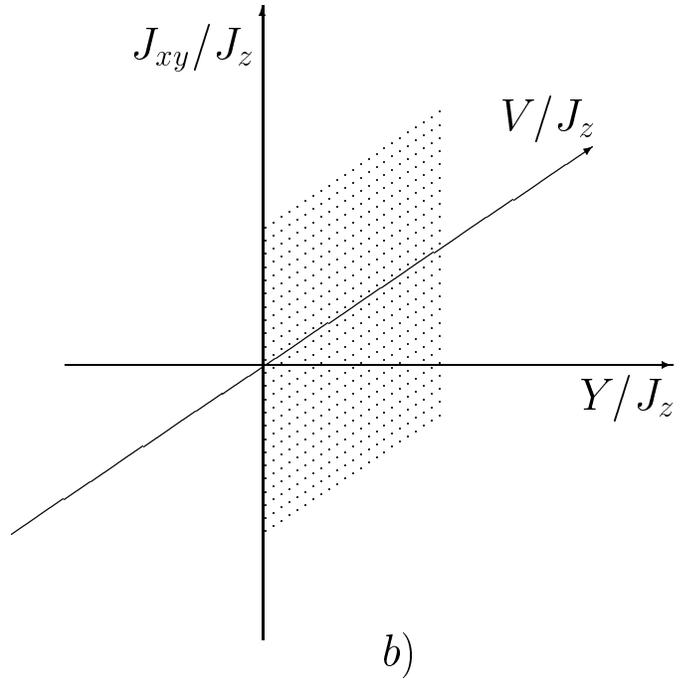
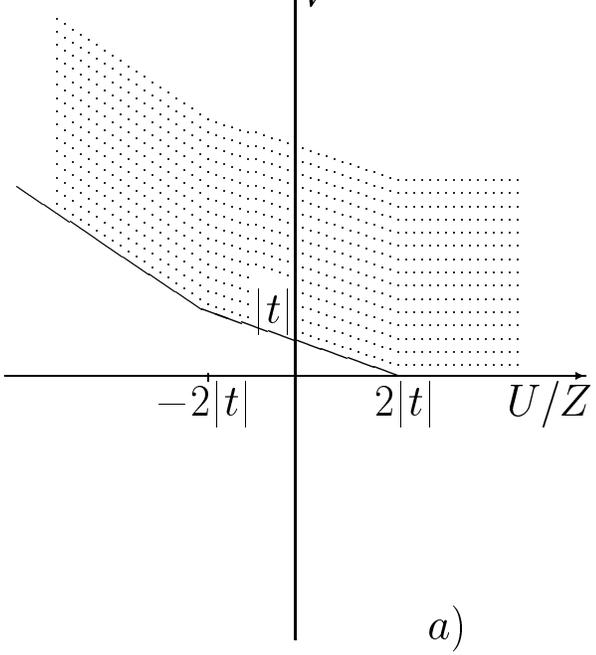
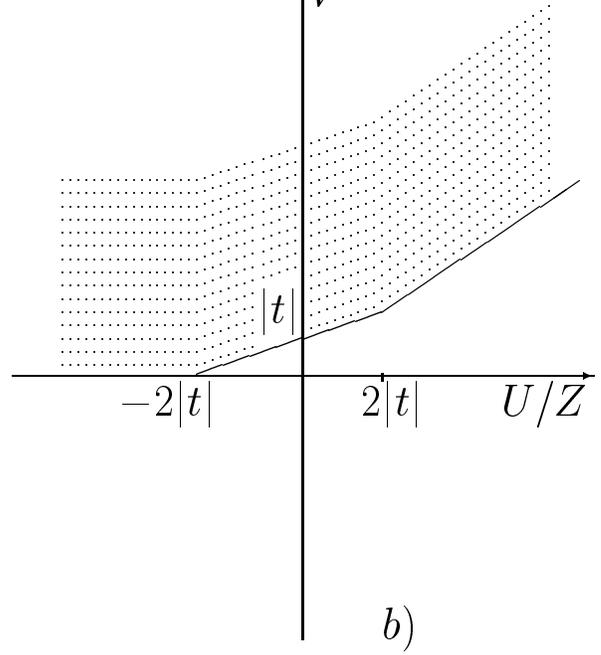
\end{document}